\documentclass[12pt,preprint]{article}
\usepackage{amssymb}

\usepackage{amsmath}
\usepackage{graphics}
\usepackage{epsfig}
\usepackage{bbm}

\newcommand{\eqb}{\begin{equation}}
\newcommand{\eqe}{\end{equation}}
\newcommand{\dmb}{\begin{displaymath}}
\newcommand{\dme}{\end{displaymath}}

\newcommand{\eab}{\begin{eqnarray}}
\newcommand{\eae}{\end{eqnarray}}

\newcommand{\e}{\mbox{e}}
\newcommand{\be}{\begin{equation}}
\newcommand{\ee}{\end{equation}}

\begin{document}

\begin{titlepage}
\begin{flushright} 
HD-THEP-07-4
\end{flushright}
\vspace{0.6cm}

\begin{center}
\Large{Yang-Mills thermodynamics at low temperature}

\vspace{1.5cm}

\large{Ralf Hofmann}

\end{center}
\vspace{1.5cm} 

\begin{center}
Institut f\"ur Theoretische Physik\\ 
Universit\"at Heidelberg\\ 
Philosophenweg 16\\ 
69120 Heidelberg, Germany
\end{center}
\vspace{1.5cm}
\begin{abstract}

\end{abstract} 

For the 
confining phase of SU(2) Yang-Mills thermodynamics 
we show that the asymptotic series representing 
the pressure is Borel summable for negative (unphysical) 
values of a suitably defined 
coupling constant. The inverse Borel transform 
is meromorphic except for a branch cut 
along the positive-real axis. The physical pressure is 
precisely nil at vanishing and acquires a 
small imaginary admixture at small temperature the latter 
indicating violations of thermal equilibrium (turbulences).

\end{titlepage}

\section{Introduction and miniature review}

To understand the (thermo-)dynamics of strongly 
coupled Yang-Mills theories is of importance 
both in view of an unearthing of their interesting mathematical 
structures and an exploitation of these theories 
in physical applications. 

As discussed in \cite{Hofmann2005_1,Hofmann2005_2}, SU(2) and SU(3) Yang-Mills 
theories occur in three phases. The deconfining, high-temperature 
phase, see also \cite{Hofmann2005D_1,Hofmann2005D_2}, possesses a nontrivial ground state which is composed 
of interacting (anti)calorons of unit 
topological charge modulus. To analytically access this 
highly complex dynamics one is lead to perform a spatial 
coarse-graining leading to an effective theory subject to a 
maximal resolution $|\phi|$. Here $\phi$ denotes an emergent, inert, 
adjoint, spatially homogeneous, and BPS saturated scalar field 
\cite{Hofmann2005_1,Hofmann2005_2,HerbstHofmann2004}. The loop expansion of thermodynamical 
quantities, carried by coarse-grained, topologically trivial 
modes of potential resolution smaller than $|\phi|$, converges 
very rapidly \cite{HofmannSept2006}: There is infrared stability due 
to the emergence of mass (adjoint Higgs mechanism), and the action of collective quantum 
fluctuations swiftly decreases with a growing number of loops. 

For a small window of intermediate temperatures a 
preconfining phase occurs, see also \cite{Hofmann2005P}. The ground 
state of this phase is composed of condensed magnetic (anti)monopoles. 
A spatial coarse-graining uniquely generates a scale of maximal resolution 
given by the modulus of a spatially homogeneous, inert, and BPS saturated 
complex scalar field $\varphi$. The stable excitations in the 
preconfining phase are noninteracting, massive, dual gauge modes, and thus the 
loop expansion of thermodynamical quantities is trivial. (It is represented by a ground-state 
plus a one-loop contribution.) 

The objective of the present work is an investigation 
of the dependence on temperature $T$ of thermodynamical quantities in 
the confining phase of SU(2) Yang-Mills theory, for detailed discussions see also \cite{Hofmann2005C,GHS2006}, 
where all gauge modes are infinitely heavy, the ground state is a condensate of 
paired, (massless) magnetic center-vortex loops, and the propagating excitations 
are massless or massive spin-1/2 fermions. These solitons are classified according to 
their topology arising from the number $n$ of selfintersections. The latter give rise to a naked 
mass $n\Lambda$, $\Lambda$ being the Yang-Mills scale (or the mass of the stable soliton with $n=1$). 
Naively, that is, when 
neglecting the excitability of internal degrees of freedom {\sl within} a given 
soliton and when disregarding the (contact) interactions {\sl between} solitons the total pressure 
is represented in terms of an asymptotic series in powers of a dimensionless 
coupling $\lambda\equiv\exp(-\Lambda/T)$. Notice that $\lambda$ is strictly 
smaller than unity for $T\le\Lambda$\footnote{At $T\sim\Lambda$ 
a Hagedorn transition occurs. The latter goes with a condensation of the selfintersection 
points within densely packed center-vortex loops into a new ground-state: The (anti)monopole 
condensate of the preconfining phase.}.

The question then 
arises whether, as a matter of principle, sufficient information is contained in the asymptotic 
series to generate the temperature dependence of the {\sl physical} pressure\footnote{By physical we mean that this quantity 
takes into account the dressing {\sl of} and the interactions {\sl between} naked solitons.}. The main result of the present work is to 
answer this question with yes albeit subject to a surprise: Our result predicts the 
exact vanishing of the physical pressure at $T=0$ and the perceptible breakdown of thermodynamical equilibrium at a 
sufficiently large temperature. This effect is manifested in terms of an 
imaginary admixture to the real pressure. Once the temperature dependence of this real part is known 
other thermodynamical quantities can be computed by Legendre transformations. Our analysis invokes a combination 
of Borel-transformation and analytical-continuation arguments for both $\lambda$ and the Borel parameter.  

The article is organized as follows. In Sec.\,\ref{nSP} we discuss and set up an 
asymptotic series representing the pressure in the confining 
phase of an SU(2) Yang-Mills theory. Modulo an algebraic-in-$n$ uncertainty 
this series is explicitly known. Sec.\,\ref{BTPP} processes the asymptotic 
series. Namely, in Sec.\,\ref{BP} we perform a Borel transformation which recasts 
the part of the pressure arising from massive excitations into a 
linear combination of polylogarithms. Some of the analyticity structure 
of the latter is discussed subsequently. In Sec.\,\ref{IBT} we perform 
the inverse Borel transformation numerically for negative arguments of 
the polylogarithms and observe that the inverse Borel 
transform is a meromorphic function except for a branch 
cut along the positive-real axis. Motivated numerical fits to the inverse 
Borel transforms of the first few relevant polylogarithms are 
carried out in Sec.\,\ref{clag0} for negative values of the argument. Subsequently, the fits are 
continued to positive values of the real part of 
the argument. It is observed that the real parts of the 
inverse Borel transforms are continuous across the cut while the moduli 
of the imaginary parts are smaller than those of the real parts for 
sufficiently small values of $\lambda$. An interpretation of this 
result (violation of thermal equilibrium by turbulences) 
is given in Sec.\,\ref{int}. In Sec.\,\ref{ol} we summarize 
our results and point towards applications and future research.  

\section{Naive series for the pressure\label{nSP}}

The excitations in the confining phase of an SU(2) Yang-Mills 
theory are generated by phase jumps and an increase in modulus 
of a complex scalar field $\Phi$ which measures the expectation of 
the 't Hooft loop operator \cite{'tHooft}. The latter is a 
dual order parameter for confinement. The phase of $\Phi$ is given by a line integral 
of the dual (abelian) gauge field along an $S^1$ of infinite radius measuring, 
by Stokes' theorem, the magnetic flux through its minimal surface ${\cal M}_1$. The creation of a magnetic center-vortex 
loop is understood as the process of having an infinitely thin flux line and its oppositely directed partner (travelling in from 
infinity) intersect with the $S^1$ which leads to their subsequent piercing the surface 
${\cal M}_1$. This brings into existence a propagating, 
single center-vortex loop. The required energy needed is provided by a 
potential $V(\Phi)$ which is initiated by a decaying 
monopole condensate at the Hagedorn phase boundary\footnote{For $|\Phi|<\Lambda$ an 
increase of $\Phi$'s modulus leads to a {\sl decrease} of $\Phi$'s energy density. At the two minima 
$\Phi=\pm\Lambda$ energy density and pressure vanish. For more information 
see \cite{Hofmann2005_1,Hofmann2005_2,Hofmann2005C}.}. The potential 
$V(\Phi)$ forces the field $\Phi$ to change its phase 
discontinuously in units of $\pi$ (units of center flux). Collisions\footnote{Only contact interactions occur due to the complete decoupling of 
propagating gauge modes in the confining phase \cite{Hofmann2005_1,Hofmann2005_2}.} 
of single center-vortex loops lead to twisting and merging thus creating 
selfintersections. This process converts kinetic energy 
into mass. In reality the very process of thermalization proceeds via instable 
high-mass excitations: Annihilations of oppositely, 
$Z_2$ charged intersection points locally elevate the 
energy density of the field $\Phi$. Subsequently, $\Phi$ ejects this energy by 
phase jumps and an increase of its modulus thus 
recreating center-vortex loops. In the hypothetic 
case of absolutely stable excitations at arbitrarily large 
selfintersection number $n$, which once and forever were thermalized to a 
given temperature by a single relaxation process of the field $\Phi$ to one of its minima, there is no ground-state 
contribution to the pressure. Here we are concerned with 
this hypothetical situation which allows for drastic 
simplifications compared to the direct 
analytical treatment of the full physical situation and, as we will see below, captures sufficient 
information to recover the latter. 

\begin{figure}
\begin{center}
\leavevmode
\leavevmode
\vspace{4.5cm}
\includegraphics{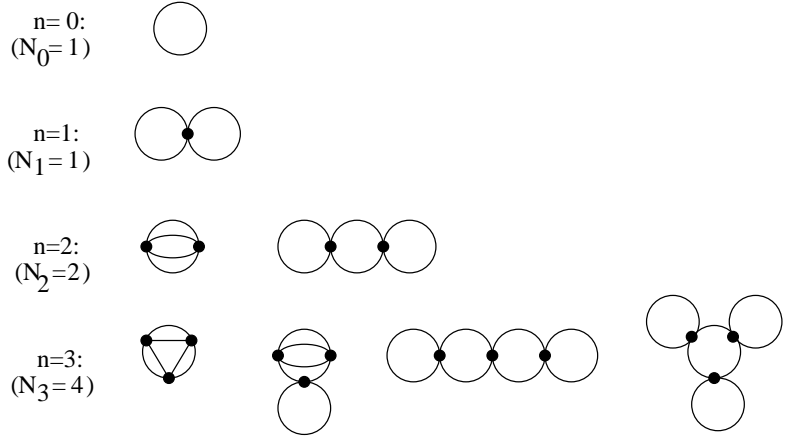}
\end{center}
\caption{A list of the soliton topologies for $n\le 3$. 
The number $N_n$ is the number of connected vacuum 
bubbles with $n$ vertices in a $\lambda\phi^4$-theory.\label{Fig-1}}      
\end{figure}

The total pressure at temperature $T$ is then represented by the 
following (asymptotic) series (sum over partial pressures of fermionic particles stemming 
from the sectors with $n$ selfintersections)
\eqb
\label{pressuredef}
P=\sum_{n=0}^\infty P_n\equiv T\sum_{n=0}^\infty M_n\int_0^\infty dp\,\frac{p^2}{2\pi^2}\,\log\left(1+\e^{-\beta\omega_n}\right)\,, 
\eqe
where $\beta\equiv \frac{1}{T}$, $\omega_n\equiv\sqrt{p^2+(n\Lambda)^2}$, $\Lambda$ the Yang-Mills scale, 
and the number $M_n$ denotes the multiplicity for solitons with $n$ selfintersections (and naked mass $n\Lambda$): 
\eqb
\label{Mn}
M_n\equiv 2\,N_n\,C_n\,.
\eqe
Here the factors of 2, of $N_n$, and of $C_n$ stand for the 
spin multiplicity, the number of distinct topologies, and the 
$Z_2$ charge multiplicity, respectively. One has (two possibilities at each intersection point)
\eqb
\label{Cn}
C_n=2^n\,.
\eqe
In Fig.\,\ref{Fig-1} we list soliton 
topologies for $n\le 3$. It is obvious that $N_n$ represents the number 
of connected vacuum bubble diagrams with $n$ vertices in a 
$\lambda\phi^4$-theory.  
Modulo algebraic-in-$n$ factors and for large $n$ the number of such diagrams in a $\lambda\phi^{2{\cal M}}$-theory 
was found to be 
\eqb
\label{numberM}
[n({\cal M}-1)]![\frac{2{\cal M}}{{\cal M}-1}]^{{\cal M}n}({\cal M}-1)^n
\eqe
by Bender and Wu \cite{BW1976}. Thus for ${\cal M}=2$ we 
have 
\eqb
\label{Knex}
N_n\sim n! 16^n\,,
\eqe
where the $\sim$ sign signals a dependence modulo a factor of the form $\sum_{l=1}^L a_l\,n^{l}\,,$ ($L<\infty$ and $a_l$ integer). The total pressure $P$ is then represented by the following 
asymptotic series:
\eab
\label{asymptSer}
P_{\tiny\mbox{as}}&=&\frac{\Lambda^4}{2\pi^2}{\hat{\beta}}^{-4}\left(\frac{7\pi^4}{180}+{\hat{\beta}}^{3}\sum_{n\ge1}M_n\int_0^\infty
dx\,x^2\,\log\left(1+\e^{-\hat{\beta}\sqrt{n^2+x^2}}\right)\right)\nonumber\\ 
&\le&\frac{\Lambda^4}{2\pi^2}{\hat{\beta}}^{-4}\left(\frac{7\pi^4}{180}+{\hat{\beta}}^{3}\sum_{n\ge1}M_n\int_0^\infty
dx\,x^2\,\e^{-\hat{\beta}\sqrt{n^2+x^2}}\right)\nonumber\\ 
&=&\frac{\Lambda^4}{2\pi^2}{\hat{\beta}}^{-4}\left(\frac{7\pi^4}{180}+{\hat{\beta}}^{2}\sum_{n\ge1}M_n\,n^2\,K_2(n\hat{\beta})\right)\nonumber\\ 
&\sim&\frac{\Lambda^4}{2\pi^2}{\hat{\beta}}^{-4}\left(\frac{7\pi^4}{180}+\sqrt{\frac{\pi}{2}}\,{\hat{\beta}}^{\frac32}
\sum_{n\ge1}M_n\,\lambda^n\,n^{\frac{3}{2}}\right)\nonumber\\ 
&\le&\frac{\Lambda^4}{2\pi^2}{\hat{\beta}}^{-4}\left(\frac{7\pi^4}{180}+\sqrt{2\pi}\,{\hat{\beta}}^{\frac32}
\sum^L_{l=0}a_l\sum_{n\ge1}(32\lambda)^n\,n!\,n^{\frac{3}{2}+l}\right)\,.
\eae
In Eq.\,(\ref{asymptSer}) we have defined $\hat{\beta}\equiv\frac{\Lambda}{T}$, $\lambda\equiv \e^{-\hat{\beta}}$, and 
$K_2$ denotes a modified Bessel function. From the $\lambda$ dependence of the sought-after 
physical pressure it should later become clear that the very concept of thermalization ceases 
to be useful for $T\nearrow\Lambda$ or for $\lambda\nearrow e^{-1}$ due to the vicinity to the 
Hagedorn transition \cite{Hofmann2005_1,Hofmann2005_2}. In Eq.\,(\ref{asymptSer}) the first $\le$ sign holds strictly for 
the linear truncation of the expansion of the logarithm about unity, and the $\sim$ sign 
indicates that terms of order $(\hat{\beta}n)^{-1}$ have been neglected in the expansion of the 
nonexponential factor in the Bessel function. The second $\le$ sign holds because 
we have used the large-$n$ expression for $N_n$ of Eq.\,(\ref{Knex}). The coefficients $a_l$, which determine the 
algebraic factor in $N_n$, are unknown at present. 

Obviously, the sum in the last line of Eq.\,(\ref{asymptSer}) diverges but exhibits the 
characteristics of an asymptotic expansion. Notice the formal similarity of the expansion in Eq.\,(\ref{asymptSer}) with 
the perturbative loop expansion of the ground-state energy in a 
$\lambda\phi^4$-theory in one dimension\footnote{In one dimension all bubble diagrams 
are finite just like the integral over thermal phase space in Eq.\,(\ref{asymptSer}) is. 
The important difference is that the coefficients of the expansion in $\lambda$ are 
alternating in sign \cite{BenderWu1969_1,BenderWu1969_2} while they are strictly positive in Eq.\,(\ref{asymptSer}).} for which Borel summability was proven, 
see \cite{KleinertBook2002} and references therein.  

\section{Borel transform and physical pressure\label{BTPP}}     

Our strategy to elute the physical pressure from the asymptotic expansion 
in  Eq.\,(\ref{asymptSer}) is to perform a Borel transformation of this series. 
Subsequently, we investigate the region of analyticity of the Borel 
transform. Only for $\lambda<0$ is the inverse Borel transform real-analytic in $\lambda$. 
However, we can analytically continue our results for $\lambda<0$ to $\lambda=\mbox{Re}\,\lambda\pm i0$\,, ($\mbox{Re}\,\lambda\ge 0$)\,, 
thus obtaining a prediction of the theory for the (perceptible) physical real part of the pressure 
at physical coupling.  

\subsection{Borel transformation\label{BP}}

The Borel transformation removes the factors of $n!$ in the coefficients of 
the power series in Eq.\,(\ref{asymptSer}):
\eqb
\label{PtoB}
\bar{P}_{\tiny\mbox{mass}}(\bar{\lambda})\equiv\sum^L_{l=0}a_l\sum_{n\ge1}{\bar{\lambda}}^n\,n!\,n^{\frac{3}{2}+l}\ 
\ \ \overset{{\tiny \mbox{Borel}}}{\longrightarrow}
\ \ \ B_{\bar{P}_{\tiny\mbox{mass}}}(\bar{\lambda})\equiv\sum^L_{l=0}a_l\sum_{n\ge1}^n {\bar{\lambda}}^n\,n^{\frac{3}{2}+l}\,,
\eqe
where $\bar{\lambda}\equiv 32\lambda$. A sum over $n$ as in Eq.\,(\ref{PtoB}) defines the polylogarithm 
$\mbox{Li}_{s}(z)$ for complex numbers $s$ and $z$ with $|z|<1$ (here $s=-\left(\frac{3}{2}+l\right)$ and $z=\bar{\lambda}$): 
\eqb
\label{polygamma}
\mbox{Li}_{-\left(\frac{3}{2}+l\right)}(\bar{\lambda})\equiv \sum_{n\ge1} {\bar{\lambda}}^n\,n^{\frac{3}{2}+l}\,.
\eqe
By analytical continuation the function $\mbox{Li}_{-\left(\frac{3}{2}+l\right)}(\bar{\lambda})$ is defined for a much larger range 
in $\bar{\lambda}$ than the definition in Eq.\,(\ref{polygamma}) seems to suggest. In any case, 
there is a branch cut for positive-real values of $\bar{\lambda}$ with $\bar{\lambda}\ge 1$. 
One has \cite{Wood92}
\eqb
\label{cut}
\mbox{Im}\,\mbox{Li}_{-\left(\frac{3}{2}+l\right)}(\bar{\lambda}\pm i0)=
\pm\pi\frac{(\log(\bar{\lambda}))^{-\left(\frac{5}{2}+l\right)}}{\Gamma(-\left(\frac{3}{2}+l\right))}\,,\ \ \ \ \ \ \ (\bar{\lambda}\ge 1)\,,
\eqe
where $\Gamma(s)\equiv\int_0^\infty dt\,\e^{-t}\,t^{s-1}$ the 
gamma function. Notice that $\Gamma(-\left(\frac{3}{2}+l\right))$ is finite and real due to the 
fractional nature of its negative argument. Notice also that there is a singularity 
$\propto (\bar{\lambda}-1)^{-\left(\frac{5}{2}+l\right)}$ at $\bar{\lambda}=1$. To the left (right) of 
$\bar{\lambda}=1$ this singularity looks like a real (imaginary), nonintegrable pole which 
follows from the limit formula \cite{Wood92}
\eqb
\label{limitLi}
\lim_{|\mu|\to 0}\mbox{Li}_s(\e^\mu)=\Gamma(1-s)(-\mu)^{s-1}\,,\ \ \ \ \ \ \ (s=-\left(\frac{3}{2}+l\right)<1)\,.
\eqe

\subsection{Inverse Borel transformation for $\bar{\lambda}<0$\label{IBT}}

The inverse Borel transform $\hat{P}_{\tiny\mbox{mass}}(\bar{\lambda})$ 
of $B_{\bar{P}_{\tiny\mbox{mass}}}(\bar{\lambda})$ is defined as
\eqb
\label{inverseBorelP}
\hat{P}_{\tiny\mbox{mass}}(\bar{\lambda})\equiv\sum_{l=0}^L a_l \hat{P}_l(\bar{\lambda})
\equiv\int_0^\infty dt\,\e^{-t}\,B_{\bar{P}_{\tiny\mbox{mass}}}(\bar{\lambda}\,t)\,,
\eqe
where
\eqb
\label{defpl}
\hat{P}_l(\bar{\lambda})\equiv\int_0^\infty dt\,\e^{-t}\,\mbox{Li}_{-\left(\frac{3}{2}+l\right)}(\bar{\lambda}\,t)\,.
\eqe
The following integral representation of $\mbox{Li}_{s}(z)$ holds for all complex $s$ and $z$ \cite{Lewin}:
\eqb
\label{intrepLi}
\mbox{Li}_s(z)=\frac{iz}{2}\int_C du\,\frac{(-z)^u}{(1+u)^s\,\sin(\pi u)}\,,
\eqe
where the path $C$ is along the imaginary axis from $-i\infty$ to $+i\infty$ with 
an indentation to the left of the origin. Inserting Eq.\,(\ref{intrepLi}) into Eq.\,(\ref{defpl}) 
for $\bar{\lambda}=-|\bar{\lambda}|\le 0$ and interchanging the order of integration, 
we have 
\eqb
\label{inverseBorel}
\hat{P}_l(\bar{\lambda})=-i\int_C du\,
\frac{(1+u)^{\frac32+l}}{1-\e^{-2\pi i\,u}}\,\e^{-\pi i\,u}\,\e^{(1+u)\log(-\bar{\lambda})}\,\Gamma(u+2)\,.
\eqe
Since, by Stirling's 
formula\footnote{$\Gamma(z)=\sqrt{2\pi}\, z^{z-\frac12}\,\e^{-z}\,e^{H(z)}$ where $H(z)\equiv\sum_{n\ge
0}\left(\left(z+n+1/2\right)\log\left(1+\frac{1}{z+n}\right)-1\right)$ 
converges for $z\in {\bf C}_{-}$ and $\lim_{|z|\to\infty} H(z)=0$, see \cite{Freitag}.} 
the gamma function $\Gamma(u+2)$ decays exponentially fast for $u\to\pm i\,\infty$ , the integral over $u$ 
in Eq.\,(\ref{inverseBorel}) exists and defines the real-analytic \footnote{This follows from Eq.\,(\ref{defpl}) and 
the fact that $\mbox{Im}\,\left[\mbox{Li}_{-\left(\frac{3}{2}+l\right)}(z)\right]\equiv 0$ for $z\le 0$.} 
function $\hat{P}_l(\bar{\lambda})\,,$ \ ($\bar{\lambda}<0$). 

\begin{figure}
\begin{center}
\leavevmode
\leavevmode
\vspace{9.0cm}
\includegraphics{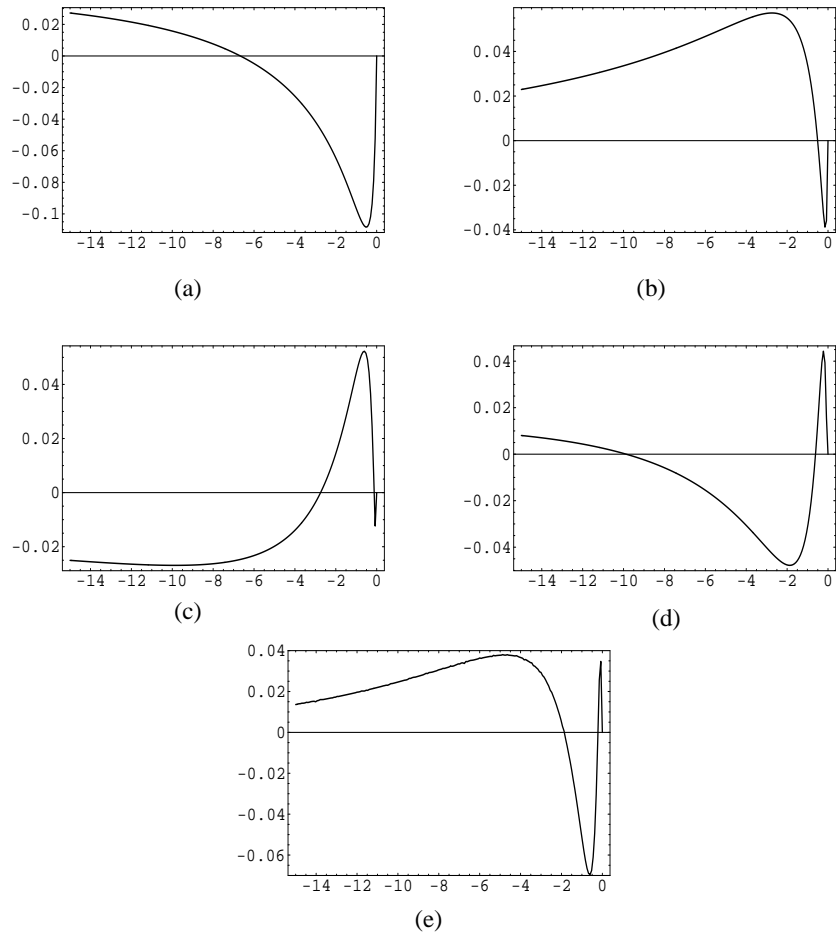}
\end{center}
\caption{The functions $\hat{P}_l(\bar{\lambda})$ for $l=0$ (a), $l=1$ (b), $l=2$ (c), $l=3$ (d), 
$l=4$ (e) and $-15\le\bar{\lambda}\le 0$. Amusingly, $l$ measures the (increasing) 
time at which snapshots are taken of an `ocean wave' crashing against 
the `beach' at $\bar{\lambda}=0$ where $\hat{P}_l\equiv 0$.\label{Fig-2}}      
\end{figure}

We have performed the inverse Borel transformation in Eq.\,(\ref{defpl}) 
numerically for $l=0,1,2,3,4$. In Fig.\,\ref{Fig-2} we have plotted 
the functions $\hat{P}_l(\bar{\lambda})\,,\ (l=0,1,2,3,4)\,,$ 
for $-15\le\bar{\lambda}\le 0$. 
Modulo powers in $u$ and for $|u|\gg 1$ the integrand in Eq.\,(\ref{inverseBorel}) roughly is of the form
\eqb
\label{Int}
c\cos(\log(-\bar{\lambda})x)\,e^{-ax}+s\sin(\log(-\bar{\lambda})x)\,e^{-ax}\,,\ \ \ \ (c,a\ \mbox{real}\,,\,a>0;\,s\ \mbox{imaginary})\,,
\eqe
where $x$ denotes the positive-real integration variable. Thus a fit 
function $\Phi_l(\bar{\lambda})$ of the following form is suggested\footnote{One has:
\eab
\label{infit}
&&\int_0^{\infty}dx\,\e^{-ax}\,\sin(\log(-\bar{\lambda})x)=\frac{a}{a^2+(\log(-\bar{\lambda}))^2}\,,\nonumber\\ 
&&\int_0^{\infty}dx\,\e^{-ax}\,\cos(\log(-\bar{\lambda})x)=\frac{\log(-\bar{\lambda})}{a^2+(\log(-\bar{\lambda}))^2}\,,\ \ \ \ (a>0)\,.
\eae
}:
\eqb
\label{fitfunct}
\Phi_l(\bar{\lambda})=\frac{\sum_{r=0}^{R_l} \alpha_{2r+1}(\log(-\gamma_{2r+1}\bar{\lambda}))^{2r+1}}{\sum_{s=0}^{S_l} 
\beta_{2s}(\log(-\delta_{2s}\bar{\lambda}))^{2s}}\,,
\eqe
where the $\gamma_{2r+1}$, $\delta_{2s}$ are positive-real, and 
$\alpha_{2r+1}$, $\beta_{2s}$ are real. The reasons for introducing higher (even and odd) 
powers of $\log(-\bar{\lambda})$ into the numerator and 
denominator of Eq.\,(\ref{fitfunct}) (not setting $\delta_{2s}=\gamma_{2r+1}\equiv 1$ and $S_l=R_l+1=1$) 
are due to the contributions 
to the integral in Eq.\,(\ref{inverseBorel}) from moderate values of $|u|$ and 
the presence of the factor $(1+u)^{3/2+l}$. We have checked that the functions displayed in Fig.\,\ref{Fig-2}, 
indeed, fall off logarithmically slowly for $-\bar{\lambda}\to\infty$. Except for the branch cut along the positive-real axis 
$\Phi_l$ is a meromorphic function of $\bar{\lambda}$. As such it can be continued arbitrarily close to 
the cut. 

\subsection{The case $\bar{\lambda}\ge 0$\label{lag0}\label{clag0}}

\begin{figure}
\begin{center}
\leavevmode
\leavevmode
\vspace{4.0cm}
\includegraphics{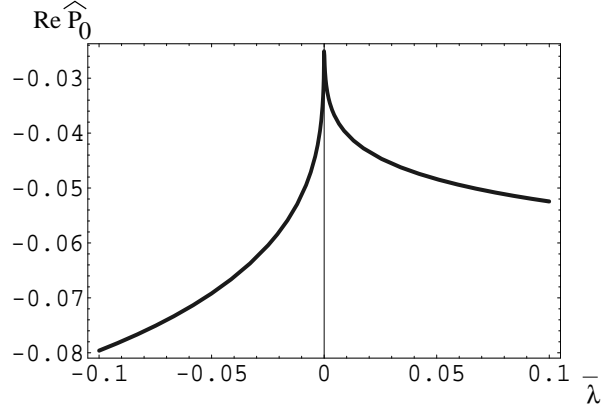}
\end{center}
\caption{The behavior of the 
function $\mbox{Re}\,\hat{P}_0(\bar{\lambda})$ in the vicinity of $\bar{\lambda}=0$.\label{Fig-E}}      
\end{figure}

\begin{figure}
\begin{center}
\leavevmode
\leavevmode
\vspace{9.0cm}
\includegraphics{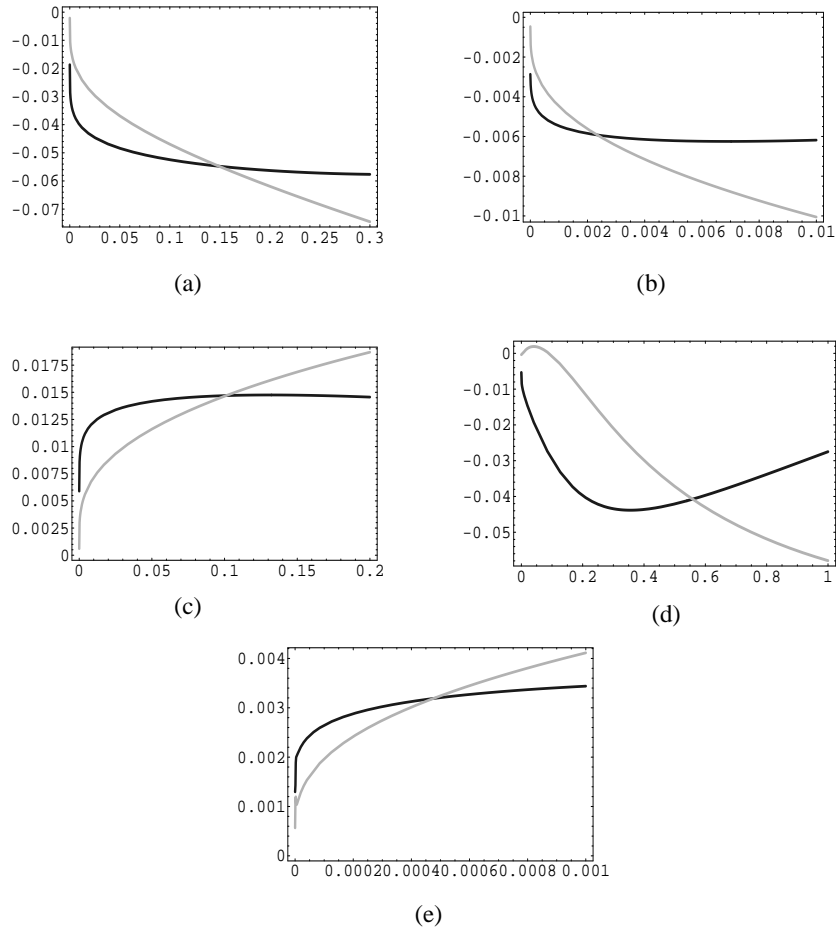}
\end{center}
\caption{The real (black) and the imaginary (grey) part of the functions $\hat{P}_l(\bar{\lambda})$ for $l=0$ (a), $l=1$ (b), $l=2$ (c), $l=3$ (d), 
$l=4$ (e) and positive $\bar{\lambda}$.\label{Fig-3}}      
\end{figure}

The functions $\hat{P}_l(\bar{\lambda})$ exhibit a branch cut along the positive-real axis. This is 
because of their dependence on powers of $\log(-\bar{\lambda})$ as suggested 
by Eq.\,(\ref{inverseBorel}). At $\bar{\lambda}=0$ there is 
a cusp in $\mbox{Re}\,\hat{P}_l(\bar{\lambda})$ with $\mbox{Re}\,\hat{P}_l(\bar{\lambda}=0)=0$, see 
Fig.\,\ref{Fig-E}. The slope of $\mbox{Re}\,\hat{P}_l(\bar{\lambda})$ at $\bar{\lambda}=\mp 0$ 
is $\pm\,\infty$.
Excellent fits reveal the following expressions for $\Phi_l\,,\ \ (l=0,1,2,3,4,5)$:
\eab
\label{fitsnum}
\Phi_0(\bar{\lambda})&=&0.0570\frac{\log(-0.154\bar{\lambda})}{1+0.220(\log(-0.494\bar{\lambda}))^2}\,,\nonumber\\ 
\Phi_1(\bar{\lambda})&=&\frac{0.0212\log(-10.2\bar{\lambda})+0.00142(\log(-0.109\bar{\lambda}))^3}
{1+0.128(\log(-1.09\bar{\lambda}))^2+
0.0544(\log(-0.886\bar{\lambda}))^4}\,,\nonumber\\ 
\Phi_3(\bar{\lambda})&=&-0.0544\frac{\log(-0.367\bar{\lambda})}{1+0.641(\log(-0.513\bar{\lambda}))^2}\,,\nonumber\\ 
\Phi_4(\bar{\lambda})&=&\frac{-0.0722\log(-1.66\bar{\lambda})+0.00780(\log(-1.95\bar{\lambda}))^3}
{1+0.212(\log(-1.90\bar{\lambda}))^2+
0.0864(\log(-1.06\bar{\lambda}))^4}\,,\nonumber\\ 
\Phi_5(\bar{\lambda})&=&\frac{14.7\log(-0.505\bar{\lambda})-1.01(\log(-0.917\bar{\lambda}))^3-0.0172(\log(-0.0568\bar{\lambda}))^5}
{1+5.84(\log(-0.0156\bar{\lambda}))^2+
0.925(\log(-10.3\bar{\lambda}))^4+1.68(\log(-0.641\bar{\lambda}))^6}\,.\nonumber\\  
\eae
Fig.\,\ref{Fig-3} shows for $\bar\lambda\ge 0$ plots of the $\mbox{Re}\,\Phi_l$, which are 
continuous across the cut, and $\mbox{Im}\,\Phi_l$, whose respective signs are chosen such as to minimize 
the value of $\bar{\lambda}>0$ where they first intersect with $\mbox{Re}\,\Phi_l$ 
(approach to the cut as $\bar\lambda=\mbox{Re}\,\bar\lambda\pm i\,0$). 
Notice that $\mbox{Re}\,\Phi_l$ and $\mbox{Im}\,\Phi_l$ vanish precisely at $\bar{\lambda}=0$. 
Notice also that for $\bar{\lambda}\ge 0$ and in the vicinity of $\bar{\lambda}=0$ the 
modulus of $\mbox{Re}\,\Phi_l$ grows much more rapidly than the modulus of $\mbox{Im}\,\Phi_l$.

\subsection{Interpretation of the results of Sec.\,\ref{lag0}\label{int}}

Due to the meromorphic nature of the functions $\Phi_l$ away from the cut along the positive-real axis 
and the fact that $\mbox{Re}\,\Phi_l$ is continuous across this cut we have no choice but to 
regard, up to a modification involving finitely many terms\footnote{We have used the large-$n$ expressions for the 
coefficients $N_n$. Precise predictions would have to subtract the first few terms from 
the expression in Eq.\,(\ref{massivePress}) (evaluated with large-$n$ coefficients) and add terms with 
realistic multiplicities.}, the quantity
\eqb
\label{massivePress}
P_{\tiny\mbox{mass}}(\hat{\beta})\equiv\frac{\Lambda^4}{\sqrt{2}\,\pi^{\frac32}\,{\hat{\beta}}^{\frac52}}
\,\mbox{Re}\,\sum^L_{l=0}a_l\,\hat{P}_l\left(\bar{\lambda}(\hat{\beta})\right)
\eqe
as a unique prediction for the pressure exerted by the massive 
modes in the confining phase of an SU(2) Yang-Mills theory, compare with Eq.\,(\ref{asymptSer}). 
We expect this pressure to be positive and rapidly growing for $\bar{\lambda}>0$, 
see \cite{GHS2006} where the asymptotic expansion of Eq.\,(\ref{asymptSer}) was used to predict the temperature dependence 
of the pressure for an `electron' gas. For accurate numerical predictions the integer 
coefficients $a_l$ would have to be known. In a $\lambda\phi^4$-theory the algebraic 
dependence of the number of connected vacuum bubble diagrams on $n$ should be 
extractable from a numerical analysis of the ground-state energy of an anharmonic oscillator, 
see \cite{BenderWu1969_1,BenderWu1969_2}. 

It is important to realize that at $\hat{\beta}=\infty$ (or $T=0$ or 
$\bar{\lambda}=0$) we have $P_{\tiny\mbox{mass}}(\hat{\beta}=\infty)=0$. That is, the contribution to the 
cosmological constant arising from an SU(2) Yang-Mills theory at zero temperature is {\sl nil}. Thus 
the according `tree-level' result \cite{Hofmann2005_1,Hofmann2005_2}, which relates to the zero of the 
effective potential for the vortex condensate ($\propto$ the 't
Hooft-loop expectation \cite{Hofmann2005_1,Hofmann2005_2}) 
survives the integration of fluctuations as performed in the present work. 

But what about the imaginary part in $P_{\tiny\mbox{mass}}$? As is readily observed 
from Fig.\,\ref{Fig-3} (and is easily proven analytically for all $l$) the modulus of 
$\mbox{Re}\,\Phi_l(\bar{\lambda})$ grows much faster than $|\mbox{Im}\,\Phi_l(\bar{\lambda})|$ with 
increasing $\bar{\lambda}$ starting at $\bar{\lambda}=0$. Thus for a certain range of 
small values\footnote{For example, $\bar{\lambda}=0.1;\,.01;\,0.0004$ corresponds to\\  
$\hat{\beta}^{-1}=\frac{T}{\Lambda}=-\left(\log\left(\frac{\bar{\lambda}}{32)}\right)\right)^{-1}=0.173;\,0.124;\,0.0886\,,$ 
respectively.} $\hat{\beta}^{-1}$ the pressure $P_{\tiny\mbox{mass}}(\hat{\beta})$ is dominated by the real part. 

The presence of a growing imaginary contamination signals an increasing deviation of the system 
from a thermodynamical behavior\footnote{No well-defined partition function directly generates 
an imaginary part for the pressure.}. Namely, we interprete the
occurrence of a sign-indefinite imaginary part\footnote{Imaginary
  contributions to the pressure lead to localized exponential built-up and 
collapse of energy density about the equilibrium situation, that is, 
to turbulences.} as an indication for the violation of a basic 
thermodynamical property: Spatial homogeneity. That is, the growth of the ratio 
$R\equiv\mbox{Im}\,P_{\tiny\mbox{mass}}/\mbox{Re}\,P_{\tiny\mbox{mass}}$ with increasing temperature 
is a measure for the increasing importance of turbulence-like phenomena in the plasma. 
At $R=1$ the thermodynamical description of the system fails badly; the system then is highly `nervous' 
and close to the Hagedorn transition. 

We believe that the so-predicted occurrence of sizably nonthermal behavior at sufficiently large 
temperature is the reason for the failure of stabilization of the magnetically confined plasma in 
tokamak experiments. This presumes that the description of the electron and its neutrino and their 
quantum mechanics \cite{PSA2005} is based on a pure SU(2) Yang-Mills
theory of scale $\Lambda=m_e=511\,$keV, see 
\cite{Hofmann2005_1,Hofmann2005_2,GHS2006}. 
Notice that the electron and its neutrino are the only stable excitations in the 
confining phase of this theory, compare with Fig.\,\ref{Fig-1}. The occurrence of poorly understood 
microturbulences and internal transport barriers in tokamaks 
was reported for $\frac{T}{\Lambda}\sim 0.05\,...\,0.09$, see JET's results and ITER's design report. 

\section{Summary and Outlook\label{ol}} 

In the present work we have performed an in-principle calculation of the pressure
\footnote{Other thermodynamical quantities are derivable from the real part of the pressure by Legendre transformations.} 
of an SU(2) Yang-Mills theory at low temperature. In particular, we have shown that the pressure vanishes precisely at 
zero temperature. We have also demonstrated that nonthermal effects start to become sizable at a certain critical temperature $T_*$. 
Numerical predictions for $T_*$ and the dependence of the pressure on temperature 
need input about the precise algebraic dependence of the number of connected vacuum bubbles on the number $n\gg 1$ of 
vertices in a $\lambda\phi^4$-theory. Also, one would need to know these 
numbers\footnote{They are known for $n\le 6$ \cite{KleinertBook2002}.} for the first few $n$. We hope that this information 
will be available soon.  

The tool employed to derive the above results is to show Borel 
summability of an asymptotic-series representation of 
the pressure at negative values of a coupling 
constant $\bar{\lambda}$ suitably defined 
for $T\ge 0$. The so-obtained functional dependence is meromorphic in the 
entire complex $\bar{\lambda}$-plane except for a branch cut along the positive-real axis. 
The crucial observation is that the real part of this function 
is continuous across the cut and that its modulus grows faster than that 
of the imaginary part for a certain range of increasing $\bar{\lambda}\ge 0$. The presence of the 
imaginary contamination signals a deviation from thermal equilibrium 
by local violations of spatial homogeneity (plasma turbulences). We believe that our results are of relevance 
in addressing the observed but poorly understood microturbulences and internal transport barriers ocurring 
in terrestial fusion experiments with magnetic plasma confinement.   
  
The generalization of our SU(2) results to SU(3) is trivial since for SU(3) the number of 
vortex-loop species at a given time is just twice that of the SU(2) case. 

\section*{Acknowledgments}

The author would like to acknowledge Gerald Dunne's, Holger Gies', and Werner Wetzel's help 
with the literature and useful conversations with Jan Pawlowski and Nucu Stamatescu. 
I am grateful for discussions with Francesco Giacosa and Markus Schwarz.

\baselineskip25pt

\end{document}